\documentclass[prd,twocolumn,superscriptaddress,altaffilletter,showpacs]{revtex4}
\usepackage{amssymb,amsmath}
\usepackage{comment}
\usepackage{graphicx}
\usepackage{graphicx} % Required for inserting images
\usepackage{amsmath,amssymb}
\usepackage{amsfonts}
\usepackage{tikz}
\usetikzlibrary{decorations.pathmorphing}
\usepackage{nccmath}

\begin{document}

\title{Nonsingular Schwarzschild-de Sitter black holes in finite conformal quantum gravity}

\author{Diego A. Martínez-Valera}
\email{diego.martinezval@alumno.buap.mx}
\affiliation{Instituto de F\'{\i}sica, Benem\'erita Universidad Aut\'onoma de Puebla,\\
Apartado Postal J-48, 72570, Puebla, Puebla, Mexico}

\date{\today }

\begin{abstract}
In this work, we prove that the classical Schwarzschild-de Sitter spacetime is an exact solution of a class of weakly non-local, UV finite conformal quantum gravity theories, without the necessity of including a cosmological constant term in the action, thus associating the effective cosmological constant $\Lambda$ appearing in the metric with the coupling constants of the quantum gravity theory. Furthermore, exploiting the inherent conformal symmetry of the theory, we take advantage of the natural enlargement of the exact solutions to motivate the construction of a regular spacetime via conformal rescaling of the Schwarzschild-de Sitter spacetime.  Moreover, we ensure the spacetime completeness by investigating the regularity of the curvature invariants and the geodesic completeness of conformally/non-conformally coupled massive and massless particles. We also study the global causal structure by explicitly constructing the Penrose diagram of the regular spacetime. Furthermore, as a result of the spacetime completeness analysis, we generalize the range of conformal factors that generate regular spacetimes, by considering the $N$ parameter of the conformal factor as a real parameter with a lower bound, and not only a positive integer, as constrained in previous studies on regular Schwarzschild/Kerr black holes. Thus, the present analysis broadens the range of solutions of the finite conformal quantum theory and opens the window to more precise observational tests of the theory using astrophysical data, by considering the accelerated expansion of the universe.
\vskip3mm

\end{abstract}

\maketitle

%: circular equatorial orbits}

%%%%%%%%%%%%%%%%%%%%%%%%%%%%%%%%%%%%%%%%%%%%%%%%%%%%%%%%%%%%%%%%%%%%

\section{Introduction} 

Einstein's theory of General Relativity is currently one of the most successful theories to explain gravitational phenomena, with remarkable predictions like black holes and gravitational waves, whose existence has been successfully supported by observations throughout the last decade, such as those provided by the Event Horizon Telescope \cite{EHT 1, EHT 2} with the shadow images of supermassive black holes (SMBHs) hosted at the core of the M87 and Milky Way galaxies, gravitational wave detections by the LIGO-Virgo collaborations caused by the merging of two stellar-mass BHs \cite{LIGO 1, LIGO 2}, as well as observations of star motion around the center of the Milky Way \cite{Ghez 1, Ghez 2, Genzel 1, Genzel 2}, to mention a few examples. Nevertheless, the theory is not free of problems and inconsistencies. Classically, a deep core problem is the controversial emergence of physical singularities, which goes against our experience of how Nature behaves. On the other hand, in the quantum realm, the efforts made in the 20$^{\text{th}}$ century have shown that four main forces govern the interactions in the universe, and all of them but gravity can be consistently formulated in the framework of a quantum field theory.

In this regard, many attempts have been made to overcome these pathologies. To begin with, the renormalizability of the theory has been a hard problem to solve, and a remarkable proposal to address this issue was provided by Stelle in 1977 \cite{Stelle}, who introduced a higher derivative quantum theory of gravity. However, even though this new theory was found to be renormalizable, it suffers from the violation of unitarity by the presence of ghosts (states of negative norm). This problem was neatly addressed by Modesto et al. \cite{Modesto1}, introducing non-local entire functions in the action while preserving causality, inspired by similar procedures applied to gauge theories \cite{Moffat1,Moffat2,Moffat3,Cornish1,Cornish2}. As a result of such modification, they obtained a finite, ghost-free theory, which is renormalizable at one loop, and finite from two loops on.

On the other hand, the scale invariance present in the theory due to the vanishing of the beta functions, and the inherent conformal invariance of the theory, have led to a thorough study of the conformal structure of the finite quantum gravity theory, made manifestly explicit under a redefinition of the metric, where the conformal invariance is achieved with the aid of a compensator field, which has no implications on the dynamics of the physical spacetime, since any physical significance it might have is wiped away once the gauge is fixed \cite{Modesto1}. In this sense, it is evident that a kind of spontaneous symmetry breaking must occur (akin to the electroweak symmetry breaking in particle physics, cf. \cite{Bambi1} for a deeper discussion) to match the lack of conformal symmetry of the observable universe. However, in the symmetry-broken phase, we expect Nature to select a gauge corresponding to our expectations of a regular spacetime. Therefore, conformal symmetry yields the appearance of a family of conformal factors able to get rid of the other long-standing problem of Einstein gravity: the physical singularities.

Several efforts have been made regarding regular spacetimes, with exact solutions like Bardeen, Hayward, or Ayón-Beato-García spacetimes \cite{Bardeen, Hayward, Beato-García1, Beato-García2}. Furthermore, a nonsingular Schwarzschild de Sitter spacetime has already been found in the context of the two-dimensional dilaton gravity by employing maximal and minimal curvature considerations \cite{NSSdS}. However, the nonsingular spacetime solution addressed in the present work comes from considering a quantum theory of gravity with remarkable features, and the resolution of the spacetime singularity is feasible by considering the symmetries of the theory.

Thus, the regular spacetime arising from this theory has two parameters: a length scale parameter $l$, and a parameter $N$, which denotes the family of conformal transformations that yield the nonsingular spacetime. In the literature, the $N$ parameter is considered to be a positive integer value, while the $l$ parameter has been constrained taking into account astrophysical observational data \cite{Bambi3,Bambi4,Bambi5}, to be proportional to the mass parameter $M$, by considering a rotating black hole  (with $N=2$). Besides, several investigations of the spacetime properties have been performed regarding energy conditions \cite{Energy Conditions}, quasi-normal modes \cite{Quasi-normal modes}, as well as frequency shift signatures, using the Herrera-Aguilar-Nucamendi method \cite{HN} (see also \cite{Martinez1} and \cite{Banerjee}), which have served to observationally constrain the $l$ parameter, employing megamaser data of the active galactic nucleus NGC 4258 \cite{Martinez2}.

In this work, we find that the Schwarzschild-de Sitter (SdS) metric is a solution to the Finite Quantum Gravity (FQG) theory without a cosmological constant, and therefore, without setting a scale in the theory. Then, we use the conformal symmetry of the theory to find a regular version of the SdS spacetime via conformal rescaling, in analogy to the previous works on Schwarzschild and Kerr spacetimes \cite{Modesto1}. We are motivated by the works \cite{HN,Martinez1,Martinez2,Villaraos}, by considering that it is required to take into account the accelerated expansion of the universe to obtain more precise estimations of the astrophysical black hole parameters using megamaser systems, and such advantages are provided by the SdS metric.

This paper is outlined as follows. In section \ref{II} we present a brief review of the Finite Conformal Quantum Gravity theory. In section \ref{III} we prove that SdS is a solution to the theory without the necessity of including a cosmological constant, and that the effective cosmological constant $\Lambda$ appearing in the metric is associated with the coupling constants of the higher derivative terms in the potential sector. Furthermore, we construct the regular spacetime by conformally rescaling SdS by an appropriate factor, and in sections \ref{IV}-\ref{VI} we show that the spacetime completeness is ensured by studying the regularity of the curvature invariants and the geodesic completeness of conformally/non-conformally coupled massive and massless particles. We also show that the $N$ parameter in the conformal factor can be extended to be a real parameter with a lower bound, while preserving spacetime completeness. In section \ref{VII} we study the global causal structure by constructing the Penrose diagram, and finally, some remarks and a final discussion are presented in section \ref{VIII}.

\section{Finite conformal quantum gravity}
\label{II}

In this section we provide a general review of the finite, D-dimensional, weakly non-local and ghost-free theory, constructed from the following general form of the lagrangian \cite{Modesto1,Modesto4}

\begin{equation}
    \mathcal{L}_{g} = -2\kappa^{-2}_{D}\sqrt{-g}\left(R + R_{\mu\nu\rho\sigma}\gamma(\square)^{\mu\nu\rho\sigma}_{\alpha\beta\tau\delta}R^{\alpha\beta\tau\delta}+V\right), 
    \label{generalAction}
\end{equation}
where we use the signature $(-,+,+,+)$ for the metric tensor, the Riemann tensor is defined by 
\begin{equation}
    R^{\mu}_{\,\,\,\nu\rho\sigma} = \partial_{\rho}\Gamma^{\mu}_{\nu\sigma}-\partial_{\sigma}\Gamma^{\mu}_{\nu\rho} +\Gamma^{\mu}_{\tau\rho}\Gamma^{\tau}_{\nu\sigma} - \Gamma^{\mu}_{\tau\sigma}\Gamma^{\tau}_{\nu\rho},
\end{equation}
the Ricci tensor is given by $R_{\mu\nu}=R^{\rho}_{\,\,\,\mu\rho\nu}$, the Ricci scalar is $R = R^{\mu}_{\,\,\,\mu}$, and $\kappa^{2}_{D} = 32\pi G_{N}$, where $G_{N} \approx6.67\times10^{-11} \text{m}^3\text{kg}^{-1}\text{s}^{-2}$ is Newton's gravitational constant.

Additionally, an important feature of the theory is encoded in the $\gamma(\square)$ operator, which helps to get rid of the ghosts, and is a weakly non-local function of the d'Alembertian operator $\square = g^{\mu\nu} \nabla_{\mu}\nabla_{\nu}$, defined as follows
\begin{equation}
    \gamma(\square)^{\mu\nu\rho\sigma}_{\alpha\beta\tau\delta} = g^{\mu\rho}g^{\nu\sigma}g_{\alpha\tau}g_{\beta\delta}\gamma_{0}(\square)+g^{\mu\rho}g_{\alpha\tau}\delta^{\nu}_{\beta}\delta^{\sigma}_{\delta}\gamma_{2}(\square),
\end{equation}
where the form factors are given by
\begin{equation}
    \gamma_{0}(\square) = -\frac{(D-2)(e^{H_{0}}-1)+D(e^{H_{2}}-1)}{4(D-1)\square}, \label{0 form factor}
\end{equation}
\begin{equation}
    \gamma_{2}(\square) = \frac{e^{H_{2}}-1}{\square}. \label{2 form factor}
\end{equation}

Following \cite{Modesto1,Modesto4,Modesto1.1,Modesto2,Modesto3,Modesto5,Modesto6,Modesto7,Modesto8,Modesto9,UniversalFQG,Tomboulis}, we propose for the entire functions $H_{i}(\square_{\mathcal{M}})$ ($\square_{\mathcal{M}} \equiv -\square/\mathcal{M}^2$, and $\mathcal{M}$ is a mass scale of the theory) to have the following structure
\begin{equation}
    H_{i}(\square_{\mathcal{M}}) = \frac{1}{2}\left[\Gamma(0,p_{i}^2(\square_{\mathcal{M}}))+\gamma_{EM}+\ln{p_{i}^2(\square_{\mathcal{M}})}\right],
\end{equation}
where $\text{Re}(z)>0$, $\gamma_{EM}=0.577216...$ is the Euler-Mascheroni constant, $\Gamma(0, z) = \int_{z}^{\infty}t^{-1}e^{-t}dt$ is the incomplete gamma function, and $p_{i}(z)$ are polynomials of degree $\gamma+\mathcal{N}+1$, where the index $i=0,2$ indicates that in general, the polynomials $p_{i}$ are different. 

Thus, from the definitions above, we have
\begin{equation}
\begin{split}
          e^{H_{i}}-1 =&\,\,\, H_{i}(z)+H_{i}^2(z)/2+ \cdots = \frac{1}{2}\left(\Gamma(0,p_{i}^2(z))+\gamma_{EM}\right.\\
          &\left.+\ln(p_{i}^2(z))\right) + \cdots,
\end{split}
\end{equation}
and the series expansion of the incomplete Gamma function reads
\begin{equation}
    \Gamma(0,z) = -\gamma_{EM}-\ln(z)-\sum_{s=1}^{\infty}\frac{(-z)^{s}}{s\,(s!)},
\end{equation}
hence the $H_{i}$ functions reduce to a series of even powers in $\square_{\mathcal{M}}$, and for example, if we take $p_{0}(z) = p_{2}(z) = z^{\gamma+\mathcal{N}+1}$, the entire functions $H_{i}(z)$ are
\begin{equation}
\begin{split}
        H_{i}(z)& = \frac{1}{2}\left(p_{i}^2(z)-\frac{p_{i}^4(z)}{2\,2!}+\cdots\right)\\
        &= \frac{1}{2}\left(z^{2(\gamma+\mathcal{N}+1)}-\frac{z^{4(\gamma+\mathcal{N}+1)}}{4}+\cdots\right).
\end{split}
\end{equation}

Therefore, in general, the form factors read
\begin{equation}
    \gamma_{i} = \frac{e^{H_{i}}-1}{\square_{\mathcal{M}}} = \frac{1}{\square_{\mathcal{M}}}\sum_{n=0}^{\infty}\frac{\tilde{p}_{n,i}(\square_{\mathcal{M}})}{n!}, \label{form factors expression}
\end{equation}
where we have defined $\tilde{p}_{n,i} = z^n \left.\left(\frac{d^n exp(H_{i}(z))}{dz^n}\right)\right|_{z=0}$. Thus, the form factors reduce to even powers of $\square_{\mathcal{M}}$, where we omit the independent terms, \textit{i.e.} $\gamma_{i}(0) = 0$.

Additionally, to achieve finiteness in any dimension, we consider the following local potential $V$, constructed to be at least cubic in the curvature \cite{Modesto1}
\begin{equation}
\begin{split}
        V=&\sum_{j=3}^{\mathcal{N}+2}\sum_{k=3}^{j}\sum_{i} c_{k,i}^{(j)} \left(\nabla^{2(j-k)}\mathcal{R}^{k}\right)_{i}\\
        +&\sum_{j=\mathcal{N}+3}^{\tau+\mathcal{N}+1}\sum_{k=3}^{j}\sum_{i} d_{k,i}^{(j)} \left(\nabla^{2(j-k)}\mathcal{R}^{k}\right)_{i}\\
        +&\sum_{k=3}^{\tau+\mathcal{N}+2}\sum_{i} s_{k,i} \left(\nabla^{2(\tau+\mathcal{N}+2-k)}\mathcal{R}^{k}\right)_{i},
\end{split}\label{Potential}
\end{equation}
where the coefficients $c_{k,i}^{(j)}, d_{k,i}^{(j)}$ and $s_{k,i}$ are coupling constants, the $\mathcal{N}$ parameter is related to the spacetime dimension $D$ by $2\mathcal{N}+4 = D+1$ (odd dimension), $2\mathcal{N}+4 = D$ (even dimension), and $\mathcal{R}$ stands for either $R_{\mu\nu\rho\sigma}$, $R_{\mu\nu}$, or $R$. 

Furthermore, the variation of the action in Eq. \eqref{generalAction} yields the following field equations \cite{Modesto1.1}
\begin{equation}
    \begin{split}
E_{\mu\nu}=&\frac{\delta\left[\sqrt{|g|}\left(R+R_{\alpha\beta\gamma\delta}\gamma^{\alpha\beta\gamma\delta}_{\rho\sigma\tau\xi}(\square)R^{\rho\sigma\tau\xi}+V\right)\right]}{\sqrt{|g|}\delta g^{\mu\nu}}\\
=& \,G_{\mu\nu}-\frac{1}{2}g_{\mu\nu}(R\gamma_{0}(\square)R)-\frac{1}{2}g_{\mu\nu}(R_{\alpha\beta}\gamma_{2}(\square)R^{\alpha\beta})\\
&+2\frac{\delta R}{\delta g^{\mu\nu}}(\gamma_{0}(\square)R)+\frac{\delta R_{\alpha\beta}}{\delta g^{\mu\nu}}(\gamma_{2}(\square)R^{\alpha\beta})\\
&+\frac{\delta R^{\alpha\beta}}{\delta g^{\mu\nu}}(\gamma_{2}(\square)R_{\alpha\beta})+ \frac{\delta\square^{r}}{\delta g^{\mu\nu}}\left(\frac{\gamma_{0}(\square^{l})-\gamma_{0}(\square^{r})}{\square^{l}-\square^{r}}RR\right)\\
&+\frac{\delta\square^{r}}{\delta g^{\mu\nu}}\left(\frac{\gamma_{0}(\square^{l})-\gamma_{0}(\square^{r})}{\square^{l}-\square^{r}}R_{\alpha\beta}R^{\alpha\beta}\right)+\frac{\delta V
}{\delta g^{\mu\nu}} = 0,
    \end{split}\label{Original EoM}
\end{equation}
where $\square^{l,r}$ acts on the left and right arguments, respectively.

\subsection{Conformal symmetry}

It is important to remark on the scale-invariant nature of the theory at the quantum level due to the vanishing of the
beta functions (for a deeper discussion see \cite{UniversalFQG}). Furthermore, thanks to the unitarity of our quantum theory of gravity and the aforementioned property of the beta functions, the $a$-theorem in four dimensions ensures that the theory possesses a hidden conformal symmetry, which can become manifest when coupled to an external field \cite{Theorems}. Thus, following \cite{Modesto1} we can enhance the study of our theory by making the conformal invariance explicit via redefinition of the metric in terms of a compensator scalar field $\phi$ 
\begin{equation}
    g_{\mu\nu} = (\phi \,\kappa_{D})^{\frac{4}{D-2}} \hat{g}_{\mu\nu},\label{physicalmetric}
\end{equation}
where $\hat{g}_{\mu\nu}$ is the \textit{background metric}, and $g_{\mu\nu}$ is the \textit{physical metric}. It is clear that this redefinition is invariant under the conformal transformations with the adequate transformation rule for $\phi$ (presented below). Here, the scalar compensator field does not account for an external physical field or matter; rather, it is a gauge degree of freedom from the conformal symmetry. Hence, the conformal transformations act on the metric and the scalar field as follows
\begin{equation}
    \hat{g}_{\mu\nu}^* = \Omega^2\hat{g}_{\mu\nu},\,\,\,\,\,\,\,\,\,\,\,\,\phi^*=\Omega^{\frac{2-D}{2}}\phi. \label{Conformal Transformation}
\end{equation}
where $\Omega^2$ is the conformal factor that generates a conformally related spacetime $g^{*}_{\mu\nu}$.

Furthermore, since the compensator field is coupled to the background metric $\hat{g}_{\mu\nu}$, there is no fifth force associated to the field, which can be readily verified by investigating the dynamics derived from action \eqref{generalAction} with respect to the physical metric $g_{\mu\nu}$. If we compute the dynamics taking into account the background metric $\hat{g}_{\mu\nu}$, additional forces appear. This is more clearly visualized when we relate the Christoffel symbols from the two frames as follows
\begin{equation}
    \Gamma(g)_{\mu\nu}^{\sigma} =  \Gamma(\hat{g})_{\mu\nu}^{\sigma} + (\text{terms involving }\partial \ln\phi).\label{fifth force}
\end{equation}

Thus, the metric comes from a super-renormalizable/finite theory which is made conformally invariant by introducing a scalar field, closely related to the conformal factor. Therefore, the theory can be analyzed from two different frames, where the background metric $\hat{g}_{\mu\nu}$, which is multiplied by the scalar field $\phi$, gives rise to a physical metric $g_{\mu\nu}$. Hence, from the point of view of the physical metric, the content of the scalar field in the non-physical frame is not associated with any dynamical feature (in contrast to other approaches to conformally invariant theories, where $\phi$ is associated with an external massive field); rather, it is a mathematical artifact to make the hidden conformal symmetry manifestly explicit, and encodes the geometrical modifications allowing for nonsingular spacetimes.

Therefore, taking into account the subtleties remarked above, we substitute Eq. \eqref{physicalmetric} into Eq. \eqref{generalAction}, and we obtain the following Lagrangian density
\begin{equation}
\medmath{\begin{split}
        &\mathcal{L}_{g} = -2\sqrt{\hat{g}}\left[\phi^2R(\hat{g})+\frac{4(D-1)}{D-2}\hat{g}^{\mu\nu}\partial_{\mu}\phi\partial_{\nu}\phi\right]\\
        &-\left.\frac{2}{k_{D}^2}\sqrt{g}\left[R(g)\gamma_{0}(\square)R(g)+R_{\mu\nu}(g)\gamma_{2}(\square)^{\mu\nu}_{\rho\sigma}R^{\rho\sigma}(g)+V(g)\right]\right|_{\phi\hat{g}},
\end{split}}
\end{equation}
where $|_{\phi\hat{g}}$ indicates that the metric $g_{\mu\nu}$ should be replaced with $\phi\hat{g}$. In this manner, the corresponding field equations (schematically) are \cite{Modesto1} 
\begin{equation}
    \begin{split}
        &\phi^2\hat{G}_{\mu\nu} = \nabla_{\nu}\partial_{\mu}\phi^2-\hat{g}_{\mu\nu}\hat{\square}\phi^2\\
        &- 4\frac{D-1}{D-2}\left(\partial_{\mu}\phi\partial_{\nu}\phi-\frac{1}{2}\hat{g}_{\mu\nu}\hat{g}^{\alpha\beta}\partial_{\alpha}\phi\partial_{\beta}\phi\right) \\
        &-\frac{\left.\delta\left[\sqrt{|g|}\left(R+R_{\alpha\beta\gamma\delta}\gamma^{\alpha\beta\gamma\delta}_{\rho\sigma\tau\xi}(\square)R^{\rho\sigma\tau\xi}+V\right)\right|_{\phi\hat{g}}\right]}{\sqrt{|g_{_{\phi\hat{g}}}|}\delta \hat{g}^{\mu\nu}},
    \end{split}\label{metric CGEoM}
\end{equation}
for the metric $\hat{g}_{\mu\nu}$, and
\begin{equation}
    \begin{split}
        &\hat{\square}\phi = \frac{D-2}{4(D-1)}\hat{R}\phi\\
        &-\frac{\left.\delta\left[\sqrt{|g|}\left(R_{\alpha\beta\gamma\delta}\gamma^{\alpha\beta\gamma\delta}_{\rho\sigma\tau\xi}(\square)R^{\rho\sigma\tau\xi}+V\right)\right|_{\phi\hat{g}}\right]}{\sqrt{|g_{_{\phi\hat{g}}}|}\delta \phi}.
    \end{split}\label{scalar CGEoM}
\end{equation}
for the scalar field $\phi$.

For the unitary gauge $\phi=\kappa_{4}^{-1}$, any Ricci-flat spacetime solves the former field equations, given that the equations are at least linear in $R_{\mu\nu}$ \cite{Modesto1}. Therefore, in the conformal phase, the theory is conformally invariant, and we construct a family of solutions by applying the conformal transformation \eqref{Conformal Transformation}. Additionally, in order to have a scale-independent theory, we impose the following relation between $\mathcal{M}$ and the Planck mass \cite{Modesto1}
\begin{equation}
    \mathcal{M}^2 = \kappa_{D}^{-\frac{4}{D-2}}.
\end{equation}

In the following section, we are going to extend the previous statement regarding Ricci flat spacetimes, to Einstein manifolds $R_{\mu\nu} = \text{constant}\times g_{\mu\nu}$.

\section{Schwarzschild de Sitter spacetime as a solution to the theory}
\label{III}

Now, we are interested in showing that the SdS spacetime solves Eq. \eqref{Original EoM}. We know that the SdS spacetime solves the classical Einstein field equations in the vacuum ($T_{\mu\nu}=0$) with a cosmological constant $\Lambda$, and the Ricci tensor is $R_{\mu\nu}=\Lambda g_{\mu\nu}$, which implies that $G_{\mu\nu}=-\Lambda g_{\mu\nu}$, and $R=4\Lambda$.

Nevertheless, including a cosmological constant term $2\Lambda$ in \eqref{generalAction} as in the Einstein-Hilbert action implies setting a scale of the theory, thus losing the conformal symmetry. In the present study, we want to preserve the structure of the theory to construct a regular spacetime by using conformal invariance. A way to achieve this goal is by carefully selecting the coupling constants of the potential so that the theory admits the SdS spacetime as a solution, with a cosmological constant parameter arising naturally from the theory by a careful selection of the adequate terms in the potential $V(\mathcal{R})$.

We start by analyzing the field equations \eqref{Original EoM}, and the action of the operator $\square$ on the Ricci tensor and Ricci scalar of the SdS spacetime. We have that the action of the differential operator $\square$ is encoded in the form factors \eqref{0 form factor} and \eqref{2 form factor}, which reduce to a function of even powers of $\square$, with no independent terms, as shown in Eq. \eqref{form factors expression}. Thus, the action of form factors $\gamma_{i}(\square)$ on $R_{\mu\nu}$ and $R$ vanishes for the SdS spacetime, since
\begin{equation}
    R^{SdS}_{\mu\nu} = \Lambda g_{\mu\nu} \Rightarrow \square R^{SdS}_{\mu\nu} =0,
\end{equation}
and
\begin{equation}
    R^{SdS} = 4\Lambda \Rightarrow \square R^{SdS} =0.
\end{equation}
Therefore, the surviving terms of Eq. \eqref{Original EoM} are
\begin{equation}
    G_{\mu\nu}+\frac{\delta V}{\delta g^{\mu\nu}} = -\Lambda g_{\mu\nu}+ \frac{\delta V}{\delta g^{\mu\nu}}\label{SdS EoM}.
\end{equation}

Furthermore, the potential in \eqref{Potential} is constructed from combinations of $R$, $R_{\mu\nu\rho\sigma}$ and $R_{\mu\nu}$. Then, if we require the left-hand side of Eq. \eqref{Original EoM} to vanish, we must fine-tune the coupling constants $c_{k,i}^{(j)}, d_{k,i}^{(j)}$ and $s_{k,i}$ to leave only the constant terms plus the terms vanishing because they involve covariant derivatives of the curvature invariants (which are constant in SdS), so that $V=V_{0}=\text{constant}$, and then
\begin{equation}
\begin{split}
  \frac{\delta S_{potential}}{\delta g^{\mu\nu}}& = \frac{\delta \sqrt{-g}\, V_{0}}{\delta g^{\mu\nu}} +\text{vanishing terms in }\nabla^{i}\mathcal{R}\\
  &= -\frac{1}{2}g_{\mu\nu}V_{0}.  
\end{split}
    \end{equation}

By analyzing the potential from Eq. \eqref{Potential}, such terms appear when $j=k$. In the particular case $j=k=3$, some of the possible combinations of the curvature tensors in the potential \eqref{Potential} read
\begin{equation}
    R^3 = 64\Lambda^3,
\end{equation}
\begin{equation}
    RR_{\mu\nu}R^{\mu\nu} = 16\Lambda^3,
\end{equation}
\begin{equation}
    R_{\mu\nu}R^{\nu\rho}R_{\rho}^{\mu} = 4\Lambda^3,
\end{equation}
and, as we can see, they are proportional to $\Lambda^3$. Besides, we omit all the terms involving the Kretschmann invariant due to their $r-$dependence, which yields the following condition for the coupling constants associated with those terms 

\begin{equation}
    \sum_{i}c^{i}\mathcal{R}_{_{Kretschmann,i}}=0.
\end{equation}

This constraint has non-trivial effects on finiteness and super-renormalizability, and under certain circumstances, this constraint can be imposed (for further discussion on the matter, see section 2.2.1 of \cite{Modesto1.1}). Thus, we assume $-V_{0} = \tilde{c}^{(3)} \Lambda^3$, where the coupling parameter $\tilde{c}^{(3)}$ encodes the proportionality constants of the combinations of $R$ and $R_{\mu\nu}$ in the potential, as well as the information of the corresponding coupling constants. 

In this manner, Eq. \eqref{SdS EoM} reads
\begin{equation}
    -\Lambda+\frac{1}{2}\tilde{c}^{(3)}\Lambda^3 = 0,
\end{equation}
which yields the following relation between the cosmological constant and the coupling parameter
\begin{equation}
    \Lambda = \sqrt{\frac{2}{\tilde{c}^{(3)}}}, \label{Lambda-Coupling parameter}
\end{equation}
where one can readily check unit consistency by knowing that $[\tilde{c}^{(3)}] = \text{mass}^{-4}$ and $[\Lambda] = \text{mass}^{-2}$.

The above derivation relating the cosmological constant $\Lambda$ and the coupling constant is only provided for illustrative purposes. In general, we can consider more curvature terms in the potential in Eq. \eqref{Potential} which are proportional to $\Lambda$, and yield a higher degree polynomial, requiring a numerical analysis to obtain an expression analogous to \eqref{Lambda-Coupling parameter}. 

Furthermore, a concern might arise regarding the finite/renormalizability of the theory once we fine-tune the coupling constant to include only some terms (\textit{e.g.} excluding terms involving the Kretschmann invariant) in order to have SdS as a solution. However, it has been shown \cite{Modesto1} that a restricted number of operators in the potential $V$ is enough to get rid of the one loop divergences in even dimension, \textit{e.g.}, in $D=4$, two operators quartic in the Ricci tensor and scalar are sufficient to end up with all beta functions identically zero, namely
\begin{equation}
    V(\mathcal{R}) = s_{R}^{(1)}R_{\mu\nu}R^{\mu\nu}\square^{\gamma-2}R_{\alpha\beta}R^{\alpha\beta}+s_{R}^{(2)}R^{2}\square^{\gamma-2}R^2.
\end{equation}

Moreover, an explicit action involving these terms, as well as $\mathcal{R}^3$, $\mathcal{R}^4$, and $\mathcal{R}^5$ terms in the potential, renders a finite quantum theory of gravity \cite{UniversalFQG} for $\gamma>3$, thus constraining the order of the polynomial $p(z)$ to be equal to or higher than 4. 

Hence, we have proven that the SdS metric is a solution to the finite conformal quantum gravity theory, where the cosmological constant appearing in the metric functions relates to the coupling constants of the potential $V$, and we have preserved conformal invariance of the theory, UV-finiteness, and unitarity.

Additionally, when we take into account the conformal version of the theory by redefining the metric as in Eq. \eqref{physicalmetric}, we obtain the field equations in \eqref{metric CGEoM} and \eqref{scalar CGEoM}. These new field equations are satisfied by the SdS metric by choosing the unitary gauge $\phi = \kappa_{4}^{-1}$. In this case, the background metric is equivalent to the physical frame once the symmetry has been spontaneously broken by fixing the gauge. Moreover, given the conformal invariance of the theory, we can apply a conformal transformation to the metric as follows $\hat{g}^*_{\mu\nu} = \Omega^2(r)\hat{g}_{\mu\nu}$, $\phi^* = \Omega^{-1}\phi$ to obtain the desired geometrical modifications, and it can be easily verified from Eqs. \eqref{physicalmetric} and \eqref{fifth force} that after gauge fixing there is no fifth force in the spacetime associated with the metric $\hat{g}_{\mu\nu}^{*}$, and therefore it is safe to compute the dynamics in this frame. 

In this study, we focus on constructing a spacetime regular in $r=0$. To this end, we choose the conformal factor
\begin{equation}
    \Omega^2(r) = \left(1+\frac{l^2}{r^2}\right)^{2},\label{ConformalFactor}
\end{equation}
where $l$ is a length scale parameter. This selection of conformal factor, as shown below, makes the curvature invariants finite everywhere, and gives rise to a geodesically complete spacetime.

Furthermore, we can apply the conformal transformation iteratively to the metric $\hat{g}_{\mu\nu}$, which allows us to choose a more general conformal factor of the form
\begin{equation}
    S(r)= \Omega_{N}^2(r) = \left(1+\frac{l^2}{r^2}\right)^{2N},\label{General Conformal Factor}
\end{equation}
which preserves the features of the former conformal factor. In this manner, we have a whole family of spacetimes that are regular via conformal transformations.

\section{Horizons and curvature invariants of the regular SdS spacetime}
\label{IV}

Once proven that the theory accepts nonsingular black holes in an expanding universe, we analyze the spacetime completeness of the new conformally related background by thoroughly studying the regularity of the curvature invariants and the geodesic completeness. We start from the SdS metric (which plays the role of the background metric $\hat{g}_{\mu\nu}$ in our theory) in standard Schwarzschild coordinates
\begin{equation}
\begin{split}
    &ds_{SdS}^2=-f\left(r\right)dt^2+f\left(r\right)^{-1}dr^2+r^2d\Omega^2,\\
    &f(r) = 1-\frac{2M}{r}-\frac{\Lambda r^2}{3},
\end{split} \label{metric1}
\end{equation}
where $d\Omega^2=d\theta^2+\sin^2\theta\, d\varphi$. Note that $r=0$ renders a physical singularity of the former metric, and from studying the solutions to $f = 0$, we can find the other singular points (coordinate singularities). These solutions are \cite{2mass}
\begin{equation}
\begin{split}
        &r_{_{BH}} = \frac{2}{\sqrt{\Lambda}}\cos{\left(\frac{\Phi}{3}-\frac{\pi}{3}\right)},\,\,\,\,\,\,\,\,r_{_{C}} = \frac{2}{\sqrt{\Lambda}}\cos{\left(\frac{\Phi}{3}+\frac{\pi}{3}\right)},\\
        & \,\,\,\,\,\,\,\,\,\,\,\,\,\,\,\,\,\,\,\,\,\,\,\,\,\,\,\,\,\,\,\ r_{_{NP}} = -(r_{_{BH}} + r_{_{C}}) = -\frac{2}{\sqrt{\Lambda}}\cos{\left(\frac{\Phi}{3}\right)} , 
\end{split}
\end{equation}
where $\Phi = \arccos{(3M\sqrt{\Lambda})}$, and $r_{_{NP}}$ is a non-physical solution. Moreover, the black hole  $r_{_{BH}}$ and cosmological $r_{_{C}}$ horizons are bounded as follows
\begin{equation}
    2M<r_{_{BH}}<3M<\frac{1}{\sqrt{\Lambda}}<r_{_{C}}<\frac{3}{\sqrt{\Lambda}},
\end{equation}
thus the black hole horizon is bounded by the classical Schwarzschild radius and the photon sphere radius. 

Moreover, as mentioned in earlier sections, we can generate a regular spacetime by rescaling the SdS metric \eqref{metric1} as follows
\begin{equation}
    ds^{*2} = \left(1+\frac{l^2}{r^2}\right)^{2N}ds_{SdS}^2,\label{rescaled metric}
\end{equation}
and the new spacetime generated by the conformal mapping preserves the singular regions of the classical SdS spacetime.

Hence, a first step to check the regularity of this new spacetime is studying the behavior of the curvature invariants. The Ricci scalar and the Kretschmann scalar of the nonsingular spacetime in the limit $r\rightarrow 0$ tend to, respectively
\begin{equation}
    R^{*}\approx\frac{24N^2r_s}{l^{4N}}r^{4N-3},\label{Ricci approx}
\end{equation}
and 
\begin{equation}
    \begin{split}
            \mathcal{K}^{*}\approx&\frac{2r_s^2}{l^{8N}}\left\{(2N-1)^2\left[2N+(2N-1)^2+(4N+1)^2\right]\right.\\
        &\left.+(4N+1)^2\right\}r^{8N-6}.\label{Kretshmann approx}
    \end{split}
\end{equation}

\begin{figure*}[tbh]
\centering
\includegraphics[width=0.33%
\textwidth]{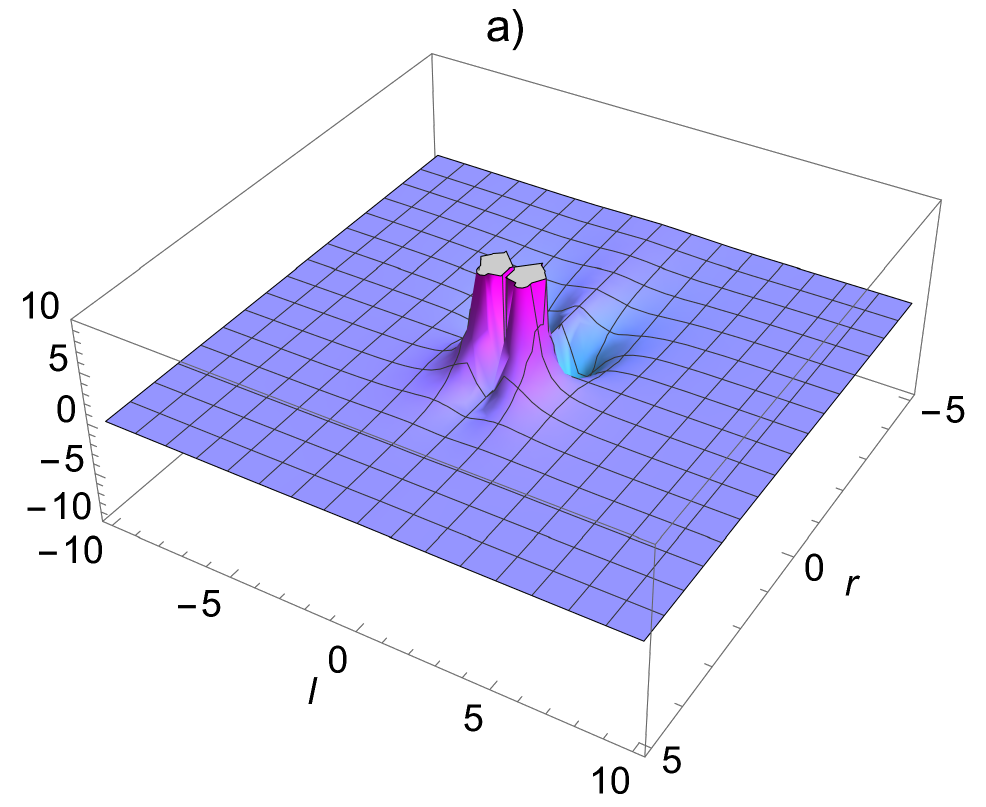}
\includegraphics[width=0.31%
\textwidth]{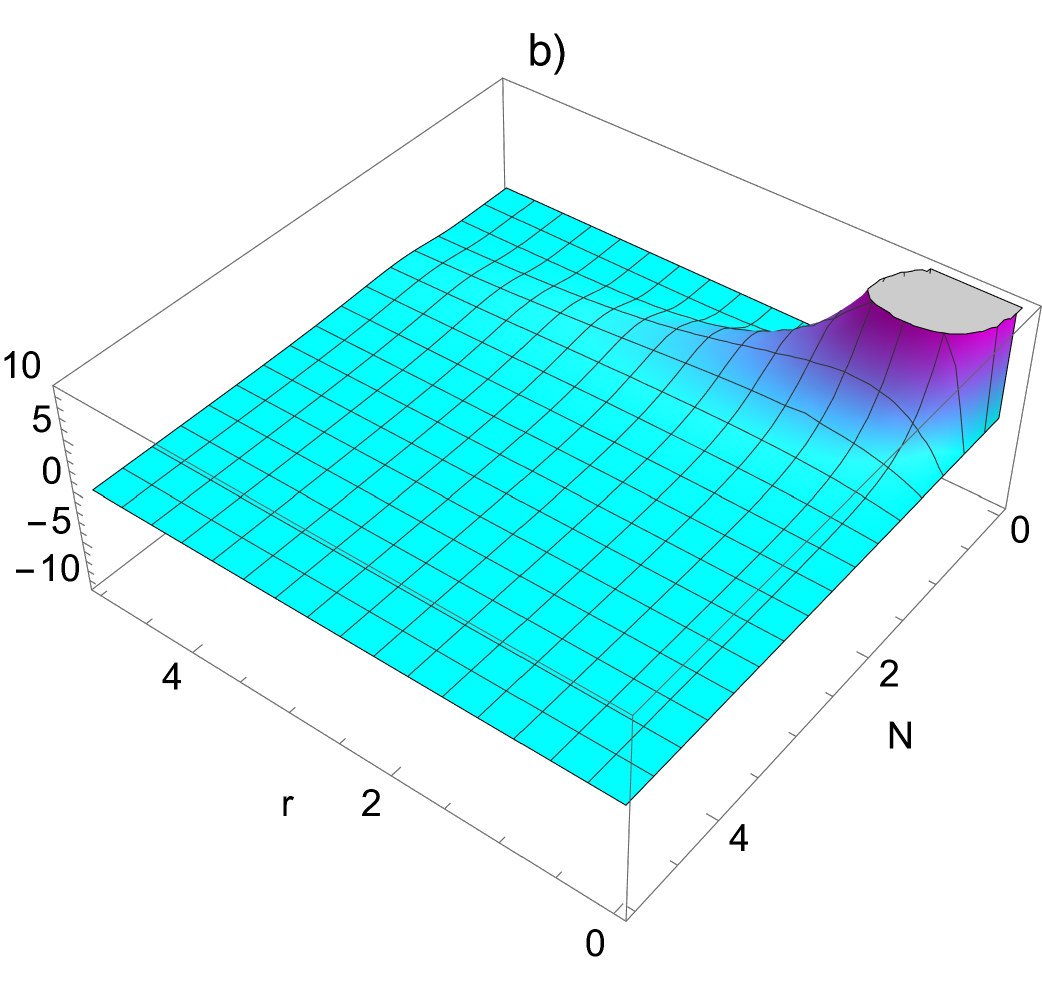}
\includegraphics[width=0.32%
\textwidth]{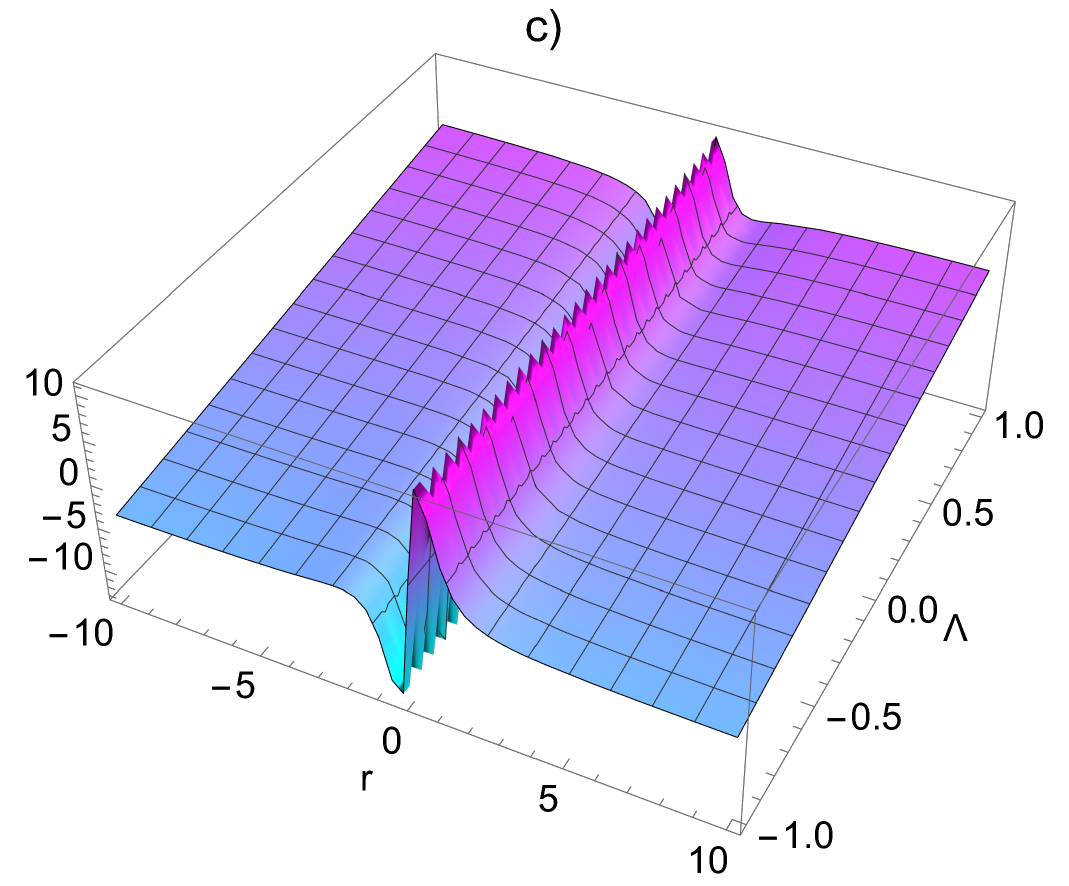}
\includegraphics[width=0.32%
\textwidth]{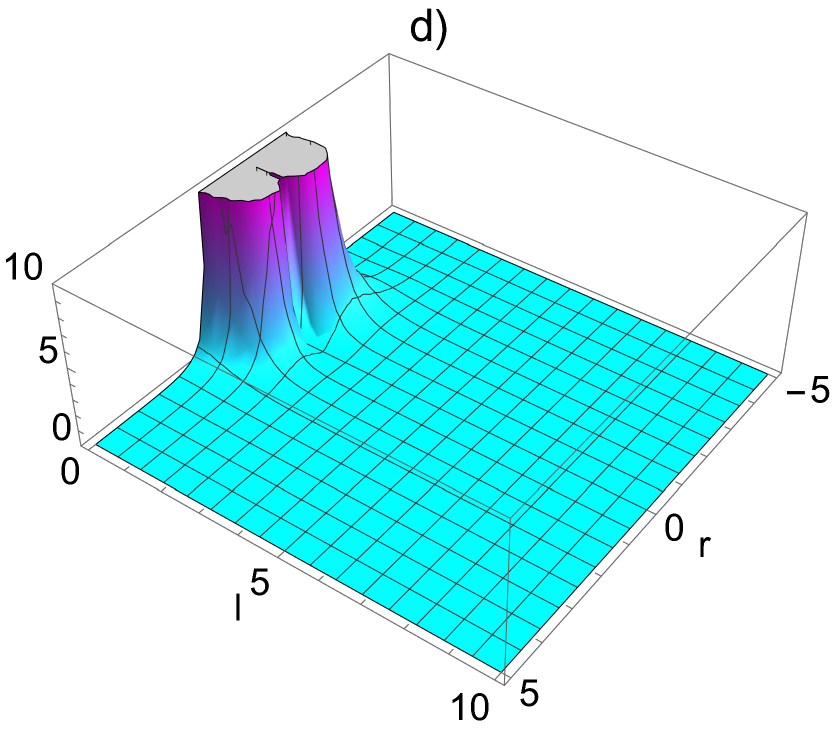}
\includegraphics[width=0.32%
\textwidth]{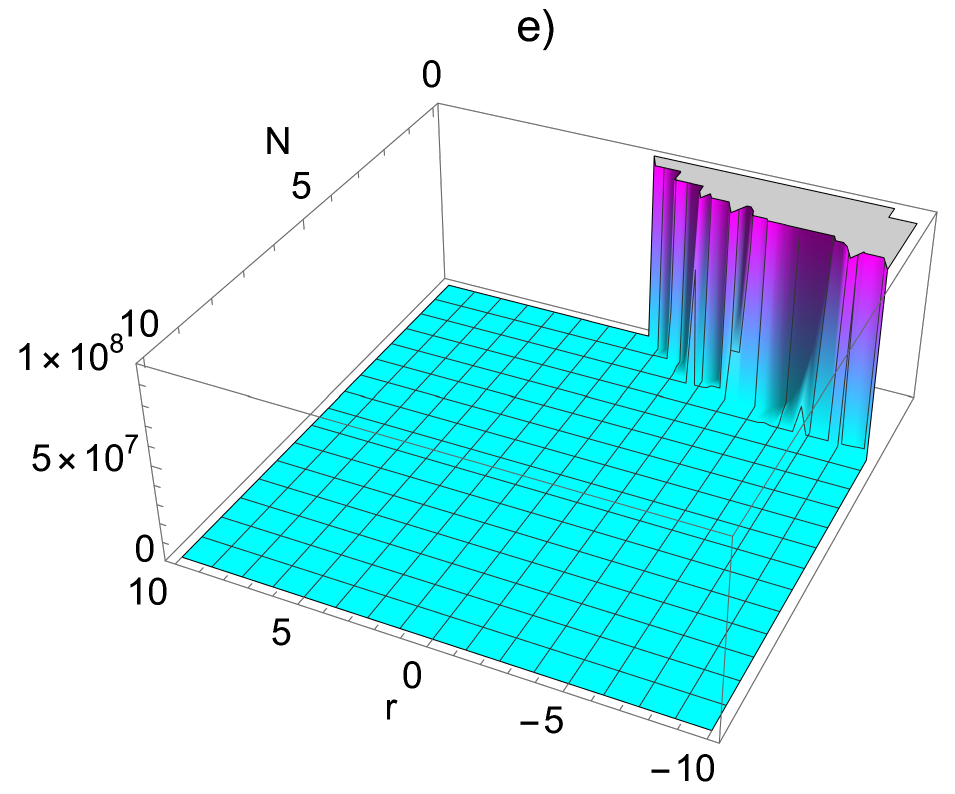}
\includegraphics[width=0.32%
\textwidth]{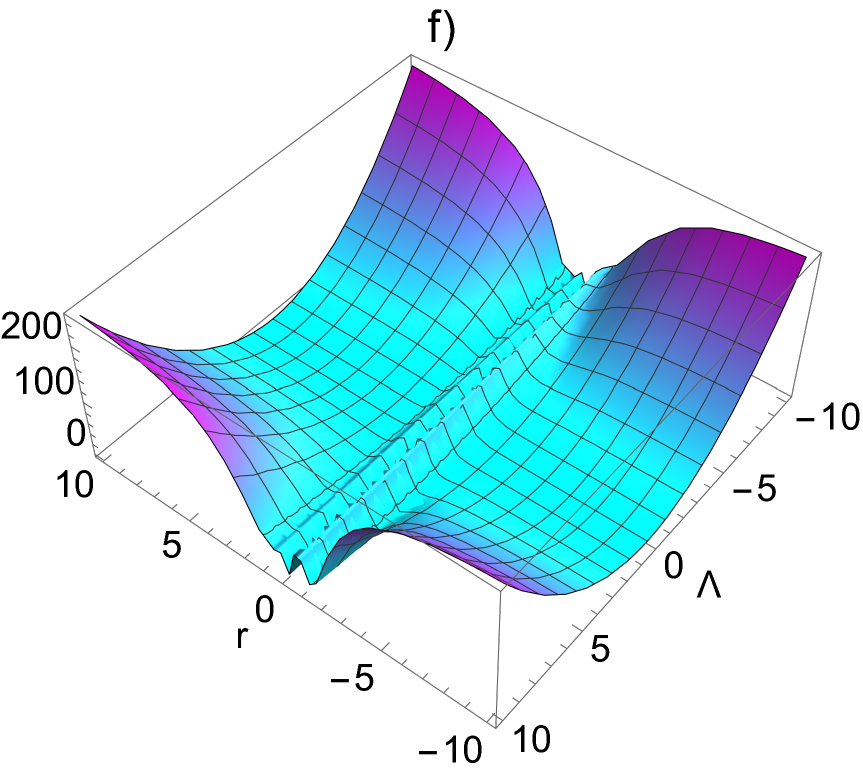}
\caption{Ricci scalar $R$ as a function of a) $r$ and $l$ with $\Lambda=0.01$, $M=1$ and $N=1$, b) $r$ and $N$ with $\Lambda=0.01$, $M=1$ and $l=1$, and c) $r$ and $\Lambda$ with with $l=M=N=1$. In the bottom plots we have the Kretschmann scalar $\mathcal{K}$ as a function of d) $r$ and $l$ with $\Lambda=0.01$, $M=1$ and $N=1$, e) $r$ and $N$ with $\Lambda=0.01$, $M=1$ and $l=1$, and f)  $r$ and $\Lambda$ with with $l=M=N=1$. We observe that both curvature invariants are regular at $r=0$ in all the plots. Nevertheless, the Kretschmann and Ricci scalars present an odd behavior for $N$ close to $0$, since its value goes to infinity as we approach small values of $N$.}
\label{Curvature Invariant Plots}
\end{figure*}

From the previous expressions, we see that the curvature invariants converge for the limit $r\rightarrow0$ as long as $N>3/4$. This fact is illustrated in Fig. \ref{Curvature Invariant Plots}, where we have plotted the Ricci scalar (panel b)) and the Kretschmann scalar (panel e)) as a function of $r$ and $N$. Additionally, Eqs. \eqref{Ricci approx} and \eqref{Kretshmann approx} show that the length scale parameter $l$ has to be different from zero (in fact, it has to be strictly positive in agreement with its physical interpretation of length) in order to ensure regularity of the curvature invariants. This statement is supported by panels a) and d) from Fig. \ref{Curvature Invariant Plots}, where we see that $R^{*}$ and $\mathcal{K}^{*}$ remain finite as long as $l>0$. Besides, panels c) and f) show that \textit{a priori}, there is no restriction on the values that the cosmological constant can acquire, since the curvature invariants are completely regular for any value of $\Lambda$.

A step further to ensure spacetime completeness is studying the geodesic completeness of such spacetime. In this regard, it is sometimes stated in the literature that geodesic completeness requires that particles never reach $r = 0$ in finite proper time or affine parameter. However, the correct criterion is more general. Geodesic completeness demands that geodesics can be extended to arbitrary values of their affine parameter, without obstruction from curvature singularities. Since curvature invariants remain finite and the geodesic can be continued smoothly through $r = 0$, the spacetime remains geodesically complete.

Hence, in accordance with \cite{Bambi1,Bambi2}, we study the geodesic completeness of three cases, namely, non-conformally coupled massive and massless probe particles, and conformally coupled test particles.

%%%%%%%%%%%%%%%%%%%%%%%%%%%%%%%%%%%%%%%%%%%%%%%%%%%%%%%%%%%%%%%
%%%%%%%%%%%%%%% Curvature Invariants 3D plot %%%%%%%%%%%%%%%%%%%%%%%%
%%%

\section{Non-conformally coupled massive and massless probe particles}

\label{V}

In this section we consider the geodesic motion of non-conformally coupled massive particles. We start by studying radial motion in spacetime \eqref{metric1}. In this regard, we use the geodesic equation
 \begin{equation}
     \hat{g}^{*}_{\mu\nu} \dot{x}^{\mu}\dot{x}^{\nu} = \kappa, \label{geodesics}
 \end{equation}
 where $\kappa = -1,0$ for massive/massless particles, and $\dot{x}^{\mu} = dx^{\mu}/d\tau$ is the 4-velocity of the particle.
 
\subsection{Massive particles}

 In the first case ($\kappa = -1$), the radial motion of massive particles is governed by the equation
 \begin{equation}
     \hat{g}^{*}_{tt}\dot{t}^2 + \hat{g}^{*}_{rr}\dot{r}^2 = -1.
 \end{equation}
 
We have a spherically symmetric spacetime, and making use of the Killing vector associated with the symmetry under time translation $\xi^{\mu} = (1,0,0,0)$, we obtain the following relations
\begin{equation}
    \frac{E}{m} = -\hat{g}^{*}_{\mu\nu}\xi^{\mu}U^{\nu} = -\hat{g}^{*}_{tt}U^{t} = -\hat{g}^{*}_{tt}\dot{t}, 
\end{equation}
and the previous equation can be written in the following way
\begin{equation}
    -\hat{g}^{*}_{tt}\hat{g}^{*}_{rr}\dot{r}^2 = \frac{E^2}{m^2} + \hat{g}^{*}_{tt} \Rightarrow  (-\hat{g}^{*}_{tt}\hat{g}^{*}_{rr})\dot{r}^2 + V_{eff} = \tilde{E}^2,
\end{equation}
which resembles a conservation-like equation, under an effective potential $V_{eff} = -\hat{g}^{*}_{tt}$, and the effective energy $\tilde{E} = \frac{E}{m}$.

\begin{figure*}[tbh]
\centering
\includegraphics[width=0.31%
\textwidth]{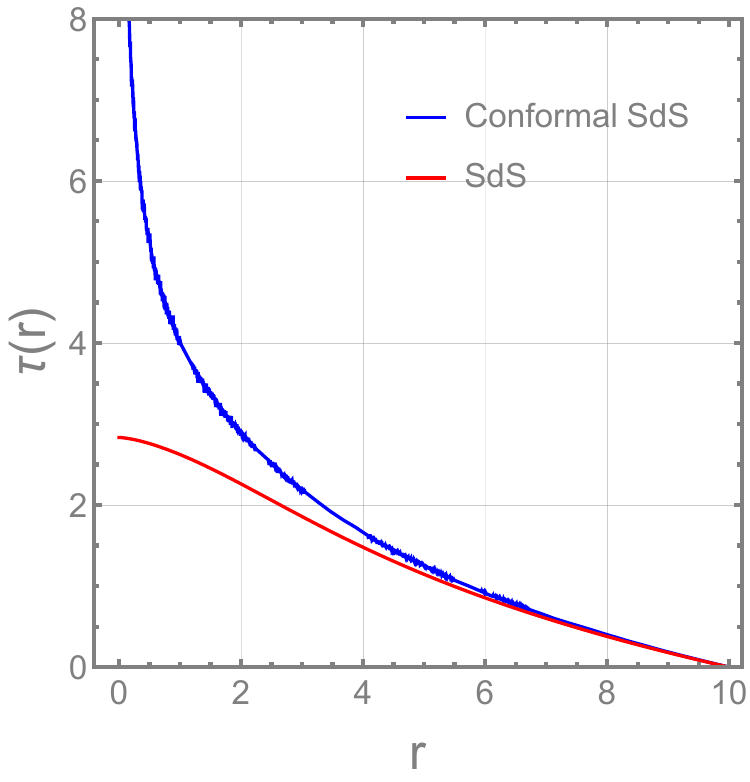}
\includegraphics[width=0.31%
\textwidth]{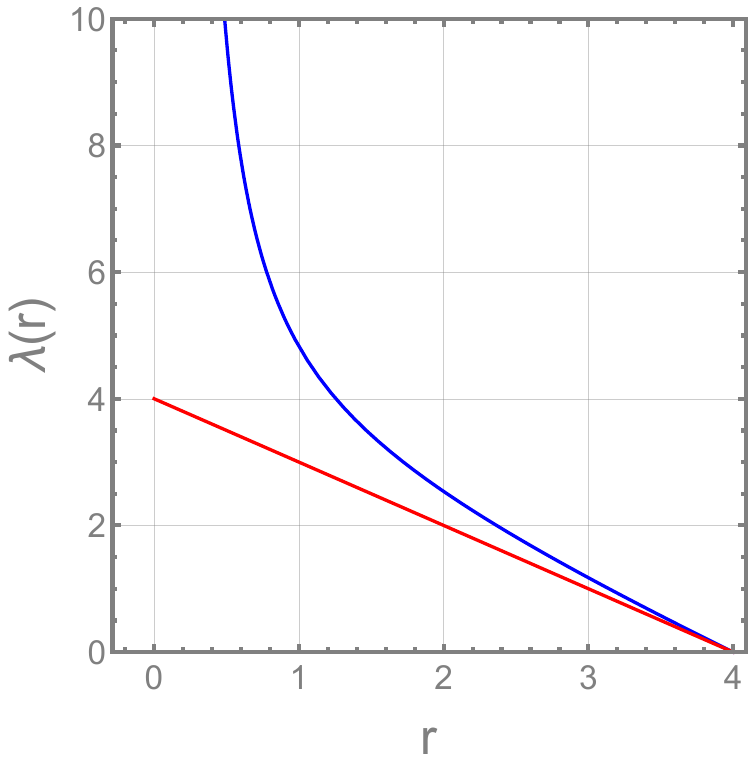}
\includegraphics[width=0.31%
\textwidth]{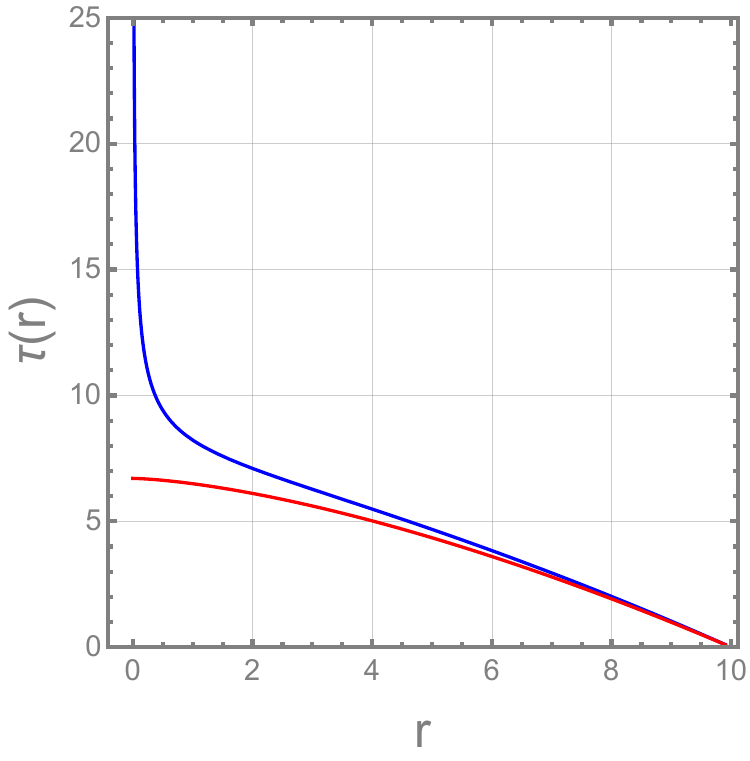}
\caption{Proper time (or affine parameter for null particles) as a function of $r$ for massive (left panel), massless (middle panel) and conformally coupled massive particle (right panel), with an effective mass $\tilde{m} = \rho\kappa_{4}$. We see that the geodesics of the spacetime generated by the conformal transformation (blue curve) requires an infinite amount of proper time to reach the former physical singularity at $r=0$, which is reached in a finite proper time in the classical SdS spacetime (red curve).}
\label{Particles geodesics}
\end{figure*}

If we rearrange the terms in this equation, using the explicit form of the conformal factor and the metric components, we obtain the following relation

\begin{equation}
    \dot{r}^2 = \frac{\tilde{E}-S(r)\left(1-\frac{2M}{r}-\frac{\Lambda r^2}{3}\right)}{S^2(r)} \label{rdot1},
\end{equation}
where $S(r) = \left(1+\frac{l^2}{r^2}\right)^{2N}$ is the conformal factor. For the sake of simplicity, we consider $N=1$ in order just to illustrate the behavior. After these considerations, we can properly study the geodesical completeness.
From Eq. \eqref{rdot1} we have
\begin{equation}
    \frac{dr}{d\tau} = \sqrt{\frac{\tilde{E}-S(r)\left(1-\frac{2M}{r}-\frac{\Lambda r^2}{3}\right)}{S^2(r)} }\label{rdot2},
\end{equation}
therefore

\begin{equation}
    \tau = -\int_{r_{0}}^{\tilde{r}} \left(\frac{S^2(r)}{\tilde{E}-S(r)\left(1-\frac{2M}{r}-\frac{\Lambda r^2}{3}\right)}\right)^{1/2} dr, \label{45}
\end{equation}
explicitly, the integral to be solved is
\begin{equation}
    \tau(\tilde{r}) = - \int_{r_{0}}^{\tilde{r}} \left(\frac{\left(1+\frac{l^2}{r^2}\right)^4}{\tilde{E}-\left(1+\frac{l^2}{r^2}\right)^2\left(1-\frac{2M}{r}-\frac{\Lambda r^2}{3}\right)}\right)^{1/2} dr, \label{nullparticlepropertime}
\end{equation}
where we assumed that the particle is falling into the black hole, hence the radial coordinate is decreasing with time $\dot{r} \leq  0$ and this is the reason why the minus sign was chosen in Eq. \eqref{45}. This integral can be addressed numerically, and the behavior in Fig. \ref{Particles geodesics} shows that an infinite proper time is required so that the particle can reach $r=0$.

\subsection{Null particles}

The geodesic equation for massless particles amounts to the case $\kappa = 0$ in \eqref{geodesics}, and it reads

\begin{equation}
    \tilde{E}^2 = -\hat{g}^{*}_{tt}\hat{g}^{*}_{rr}\dot{r}^2 = S^2(r)\,\dot{r}^2,
\end{equation}
where the overdot indicates differentiation with respect to the affine parameter $\lambda$. From this, we obtain 
\begin{equation}
    \lambda =  -\int_{r_0}^{\tilde{r}} \frac{S(r)}{\tilde{E}}dr.\label{Massless integral}
\end{equation}
This integral is straightforward and renders the following result
\begin{equation}
    \lambda = \frac{1}{\tilde{E}}\left(\frac{l^4}{3r^3}-\frac{l^4}{3r_{0}^3}+\frac{2l^2}{r}-\frac{2l^2}{r_{0}}-r+r_{0}\right),
\end{equation}
and as it can be readily verified, as $r\rightarrow 0$, the right-hand side of the latter equation diverges. Hence, the affine parameter $\lambda$ tends to infinity, as illustrated in Fig. \ref{Particles geodesics}.

\subsection{Radial null geodesics for general $N$}

Furthermore, the latter integral for a general $N$ renders the following general solution
\begin{equation}
    \lambda= \frac{l^{4N}\,_2F_1\left(\frac{1}{2}-2N,-2N;\frac{3}{2}-2N;-\frac{r^2}{l^2}\right)}{(4N-1)r^{4N-1}},
\end{equation}
where $_2F_1$ is the Hypergeometric function defined by the power series
\begin{equation}
    \sum_{n=0}^{\infty} = \frac{(a)_{n}(b)_{n}}{(c)_{n}}\frac{z^2}{n!},\,\,\,\,\,\,\,|z|<1,
\end{equation}
and the Pochhammer symbols are given by
\begin{equation}
    (s)_{n} = 
     \begin{cases}
       1 &\quad\text{if }\,\,n=0,\\
       q(q+1)\cdots(q+n-1) &\quad\text{if }\,\,n>0. \\
     \end{cases}
\end{equation}

\begin{figure}
    \centering
    \includegraphics[width=0.5%
\textwidth]{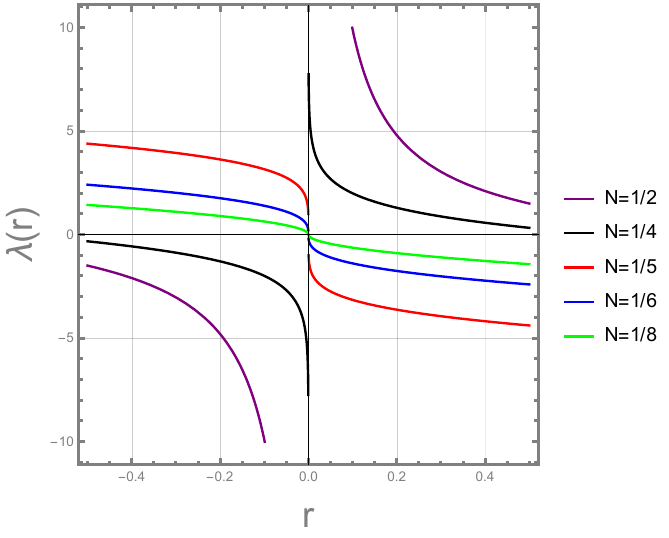}
    \caption{Affine parameter behavior for small $N$ with $l=1$, $\tilde{E}=1$. We have that the particle now approaches $r=0$, but the functions corresponding to $N<1/4$ have a domain $\mathbb{R}\backslash\{0\}$, therefore, the geodesics for $N<1/4$ are not defined for all the values of their affine parameter $\lambda$.}
    \label{SmallN}
\end{figure}

Additionally, if we take into account any real valued $N$, we see that for $N<1/4$ the behavior of the affine parameter changes as seen in Fig. \ref{SmallN}, thus, geodesic completeness does not hold for $N<1/4$. This feature is clearer if we see that as $r\rightarrow0$, the conformal factor approximates as follows
\begin{equation}
    \left(1+\frac{l^2}{r^2}\right)^{2N} \simeq  l^{4N}r^{-4N},
\end{equation}
and therefore, the integral from Eq. \eqref{Massless integral} behaves like
\begin{equation}
    \lambda(r) \simeq l^{4N}\int r^{-4N}dr,  
\end{equation}
which converges for $N>1/4$. Furthermore, from the discussion of Sec. \ref{IV}, we found that the regularity of curvature invariants is ensured for $N>3/4$, thus, a regular and complete spacetime is achieved under a conformal rescaling of the form \eqref{General Conformal Factor}, as long as $N>3/4$. 

Therefore, we can generalize the conformal factor presented in previous studies \cite{Modesto1} to consider a real-valued $N>3/4$.

\section{Conformally coupled massive probe particles}
\label{VI}
\label{coupled massive}

For the study of conformally coupled massive test particles, we start from the following conformal gravity action
 \begin{equation}
    S_{_{CG}} = - \int \sqrt{-\rho^2\phi^{*2}\hat{g}^{*}_{\mu\nu}\dot{x}^{\mu}\dot{x}^{\nu}}d\lambda.
 \end{equation}
 
 This action stands for a Weyl invariant theory with a compensator $\phi^*$, and has the associated lagrangian
\begin{equation}
    L_{_{CG}} = -\sqrt{-\rho^2\phi^{*2}\hat{g}^{*}_{\mu\nu}\dot{x}^{\mu}\dot{x}^{\nu}}. \label{Lagrangian1}
\end{equation}
which resembles the usual lagrangian of a free particle in GR, with an effective mass $\rho\phi^*$, $\rho>0$. Here we choose the unitary gauge $\phi^* = \kappa_{4}^{-1}$. Thus, we are dealing with an action for a particle with an effective mass $\tilde{m} = \rho \kappa_{4}^{-1}$. It can be readily verified that the spacetime coordinates $t, \varphi$ are cyclic in the Lagrangian \eqref{Lagrangian1}. This leads to conserved quantities associated with their respective energy and angular momentum, as used before. 

In this way, we can make use of the analogue equations to the non-conformally coupled case with a mass $\tilde{m} = \rho \kappa_{4}^{-1}$, hence we have
\begin{equation}
    \dot{r}^2 = \frac{S(r)\frac{E^2\kappa_{4}^{2}}{\rho^2}-S(r)\left(1-\frac{2M}{r}-\frac{\Lambda r^2}{3}\right)}{S^2(r)} \label{rdot}.
\end{equation}

Therefore, the equation for the affine parameter, which we have chosen to be $\lambda = \tau$ due to the proper time gauge, reads

\begin{equation}
    \tau = -\int_{r_{0}}^{\tilde{r}} \left(\frac{S^2(r)}{S(r)\frac{E^2\kappa_{4}^{2}}{\rho^2}-S(r)\left(1-\frac{2M}{r}-\frac{\Lambda r^2}{3}\right)}\right)^{1/2}dr,
\end{equation}
which for a particle at rest at infinity, $E=\rho\kappa^{-1}_{4}$, and the integral simplifies to
\begin{equation}
    \tau = -\int_{r_{0}}^{\tilde{r}} \left(\frac{S^2(r)}{\frac{2M}{r}+\frac{\Lambda r^2}{3}}\right)^{1/2}dr,
\end{equation}
where, similarly to the integral in Eq. \eqref{nullparticlepropertime}, it is required an infinite proper time to reach the point $r=0$ (see Fig. \ref{Particles geodesics}).

\section{Global causal structure}
\label{VII}
In this section, we review the general procedure to construct the Penrose diagram of a spherically symmetric black hole \cite{2mass}, and from that, we schematically indicate the construction of the diagram for the rescaled SdS spacetime.

To construct the diagram, we focus on the metric function $\mathcal{F}(r)$ and its roots $\hat{r}$. Then, we Taylor expand the metric function up to linear order as follows
\begin{equation}
    \mathcal{F}_{approx}(r) = \sigma(r-\hat{r}),\,\,\,\,\,\,\sigma = \mathcal{F}'(\hat{r})\neq 0.
\end{equation}

Then, we introduce the tortoise coordinate $r^{*}$ given by
\begin{equation}
    r^{*} = \int \mathcal{F}^{-1}(r)dr,
\end{equation}
and therefore, using $\mathcal{F}_{approx}(r)$
\begin{equation}
    r^{*}_{approx} = \int \mathcal{F}^{-1}_{approx}(r)dr = \sigma^{-1}\ln|\sigma(r-\hat{r})|,\label{r approx}
\end{equation}
which serves to introduce the advanced and retarded null coordinates $u= t+r^{*}$ and $v = t-r^{*}$. Using $u$ and $v$, the line element can be re-expressed as follows
\begin{equation}
    ds^2 = -\mathcal{F}(r)dudv + r^2d\Omega^2.\label{LineElementdudv}
\end{equation}
 
By making use of Eq. \eqref{r approx}, one can readily verify that
 \begin{equation}
     e^{\sigma r^{*}_{approx}} = e^{\sigma (u-v)/2} = \mathcal{F}_{approx}(r),
 \end{equation}
 which motivates the definition of the Kruskal coordinates $U = \mp e^{\sigma u/2}$ and $V= e^{-\sigma v/2}$. Thus,
 \begin{equation}
     dudv = -\frac{4}{\sigma^2 UV}dUdV.
 \end{equation}

 In this way, the line element \eqref{LineElementdudv} reads
 \begin{equation}
     ds^2 = \frac{4\mathcal{F}(r)}{\sigma^2 UV}dUdV + r^2d\Omega^2,\label{PD}
 \end{equation}
and it is straightforward to check that this expression is regular in $r=\hat{r}$ by writing $\mathcal{F}$ as the product of its roots. Thus, the $UV$ factor in the denominator, which includes the root we are approaching to, will cancel the one in the numerator. Moreover, we  introduce the following rescaling of the Kruskal coordinates
\begin{equation}
    \tilde{U}= \arctan{U},\,\,\,\,\,\tilde{V}=\arctan{V}, \label{compactification}
\end{equation}
which serves as a compactification of the manifold and yields the well-known conformal structure of the Penrose diagram.

 Now, to address the construction of the Penrose diagram for the metric \eqref{rescaled metric}, we note that the conformal rescaling does not affect the null geodesics, and therefore, the tortoise coordinate $r^{*}$ remains unchanged with respect to the one from the SdS spacetime
 \begin{equation}
     \begin{split}
              &r^{*} = \int\frac{rdr}{r-2M-r^3\frac{\Lambda}{3}} = \frac{3}{\Lambda}\left[\frac{r_{_{BH}}ln(r-r_{_{BH}})}{(r_{_{C}}-r_{_{BH}})(r_{_{BH}}-r_{_{NP}})}\right.\\
              &-\left.\frac{r_{_{C}}ln(r-r_{_{C}})}{(r_{_{C}}-r_{_{BH}})(r_{_{C}}-r_{_{NP}})}+\frac{r_{_{NP}}ln(r-r_{_{NP}})}{(r_{_{BH}}-r_{_{NP}})(r_{_{C}}-r_{_{NP}})}\right].
     \end{split}
 \end{equation}

Using Eq. \eqref{PD} we have
 \begin{equation}
     ds^2 = \left(\frac{4\mathcal{F}_{SdS}(r)}{\sigma^2}\right)e^{-\sigma r^{*}}dUdV + r^2d\Omega^2,
 \end{equation}
 where
 \begin{equation}
     \sigma = \left.\mathcal{F}'(r)\right|_{r=\hat{r}} = \left.(S(r))'f(r) + S(r) f'(r)\right|_{r=\hat{r}},
 \end{equation}
 where $f(r)$ is defined in Eq. \eqref{metric1}, and $\hat{r}$ can be either $r_{_{BH}}$, $r_{_{C}}$, or $r_{_{NP}}$ for the maximally extended region.

\begin{figure}[htb]
\centering
 \includegraphics[width=0.5%
\textwidth]{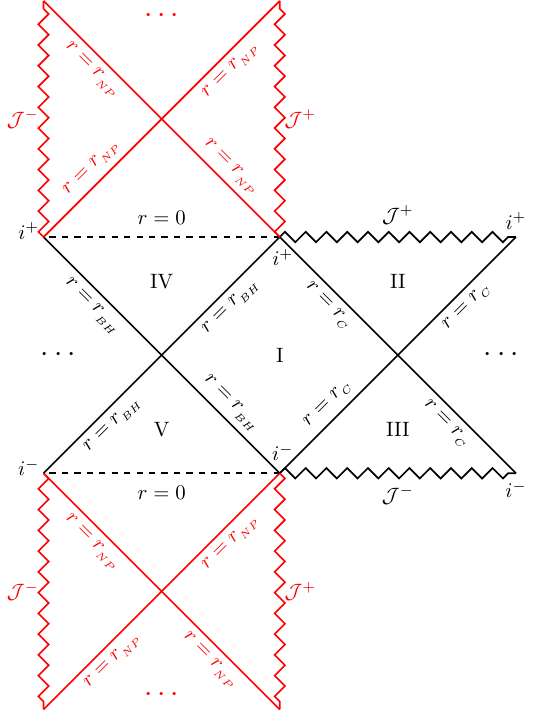}
    \caption{Penrose-Carter diagram for the singularity-free Schwarzschild spacetime. The conformal factor $S(r)$ does not change the general form of the diagram. Nevertheless, the spacetime is regular at $r=0$. The dots indicates that the displayed pattern repeats itself infinitely in those directions. The red regions are prohibited for all particles, in classical SdS due to the singularity at $r=0$, and in nonsingular SdS due to the fact that the required proper time to reach $r=0$ is infinite.}
    \label{fig:Penrose-Diagram}
\end{figure}

 Then, we apply the compactification of Eq. \eqref{compactification}, which finally renders the Penrose diagram shown in Fig. \ref{fig:Penrose-Diagram}. The diagram shows that the causal structure of the conformal equivalent spacetime resembles the SdS structure. However, the difference arises in the region $r=0$, where now it is possible to extend the region for $r\rightarrow -\infty$. Hence, we include an extended region which resembles a \textit{mirror-like universe}, corresponding to $r<0$ in the diagram and thus, taking into account the negative root $r=-(r_{_{BH}}+r_{_{C}})$ in the maximal analytic extension of the diagram. Nevertheless, as seen from the proper time plots, no particles can pass the $r=0$ region, and therefore, it is still a forbidden region.

\section{Discussion and final remarks}
\label{VIII}
In this work we have explicitly shown that the SdS spacetime is an exact solution to the FQG theory without a cosmological constant \textit{i.e.}, without setting a scale, which preserves conformal symmetry under a metric redefinition with an auxiliar scalar field. Additionally, an essential result of this work is the realization that the cosmological constant $\Lambda$ appearing in the SdS metric arises naturally from the structure of the FQG theory, specifically through the coupling constants of the higher-order curvature potential $V$. This fact eliminates the need to introduce a cosmological constant by hand, preserving the scale invariance of the theory at the classical level. In this framework, the cosmological constant emerges from a precise balancing of curvature invariants in the potential $V$. The derived relation between $ \Lambda$ and the coupling constants suggests that the cosmological constant is not fundamental but a dynamical outcome of the underlying quantum gravity structure. 

Additionally, we exploit the conformal symmetry of the theory to overcome the singularity problem at $r=0$ in SdS via conformal rescaling using an appropriate family of conformal factors. Then, we check spacetime completeness by analyzing the curvature invariants and the geodesic completeness of conformally coupled/non-coupled massive particles and null particles. The latter analysis helps us broaden the family of conformal transformations that render nonsingular spacetimes by generalizing the values of the parameter $N$ to real values with a lower bound $N>3/4$. This turns out to be very helpful because the promotion of the $N$ parameter from positive integers to real values allows us to estimate its value via Bayesian statistical fits, as previously done with the nonsingular Schwarzschild black hole \cite{Martinez2}. Furthermore, recent studies on the Bayesian statistical fit of the SdS black hole parameters have rendered remarkable results regarding the Hubble constant and have allowed for the establishment of a more direct relationship between observables and the black hole parameters, without the necessity of using special relativistic approximations to account for the expansion of the universe. This exact relation is essential when estimating sensible data to obtain the values of the black hole and spacetime parameters in modified gravity, as in this case. Furthermore, cosmological observations can be used to place constraints on the quantum gravity parameters to which $\Lambda$ is related. Thus, precise measurements of the Hubble constant, as well as the frequency shift of light emitted by particles orbiting in the vicinity of a black hole, could provide empirical input to narrow down the allowed parameter space of the potential sector of FQG theory. This connection bridges the gap between quantum gravity phenomenology and cosmological data, opening a path toward observationally testing aspects of UV-complete theories of gravity.

Further studies on this matter include a parameter estimation of the nonsingular SdS black hole, analysis of the energy conditions, quasi-normal modes, stability, thermodynamics, and black hole formation, just to mention a few.

\section*{Appendix: Curvature invariants}
The full expression of the Kretshmann scalar of the regular spacetime reads
\begin{widetext}
\begin{equation}
    \begin{split}
        &\mathcal{K}^* = \frac{\left(1+\frac{l^2}{r^2}\right)^{-4N}}{9r^6\left(l^2+r^2\right)^4} \left\{-2\left[ \left(-l^2(2N-1)+r^2\right)^2\right.\right.\left.\left(\Lambda r^3-3r+6M\right)+3r\left(l^2+r^2\right)^2\right]\left[\left(-l^2(2N-1)+r^2\right)^2\times\right.\\
        &\left(-\Lambda r^3+3r-6M\right)\left.+3r\cot^2(\theta)\left(l^2+r^2\right)^2-3r\csc^2(\theta)\right.\left.\left(l^2+r^2\right)^2\right]+\left[\left(-l^2(2N-1)+r^2\right)^2\left(-\Lambda r^3+3r-6M\right)\right.\\
        &\left.+3r\cot^2(\theta)\left(l^2+r^2\right)^2-3r\csc^2(\theta)\left(l^2+r^2\right)^2\right]^2+\left[\left(-l^2(2N-1)+r^2\right)^2\left(\Lambda r^3-3r+6M\right)\right.\left.+3r\left(l^2+r^2\right)^2\right]^2\\
        &+2\left(-l^2(2N-1)+r^2\right)^2\left[-l^2\left(-4N\left(\Lambda r^3-3r+6M\right)+2\Lambda r^3\right.\right.\left.\left.-6M\right)-2\Lambda r^5+6Mr^2\right]^2+4\left[l^4\left(-2N\left(\Lambda r^3+3r-12M\right)\right.\right.\\
        &\left.\left.+\Lambda r^3+6M\right)+2l^2r^2\left(N\left(\Lambda r^3-9r+24M\right)+\Lambda r^3+6M\right)\right.+\left.\Lambda r^7+6Mr^4\right]^2+2\left(l^4(2N-1)\left(6M-2\Lambda r^3\right)+2l^2r^2\right.\\
        &\left.\left.\left(N\left(2\Lambda r^3-12r+30M\right)+2\Lambda r^3-6M\right)+2\Lambda r^7-6Mr^4\right)^2\right\}.
    \end{split}
\end{equation}
\end{widetext}

and the Ricci scalar is
\begin{equation}
    \begin{split}
          & R^* = \frac{4\left(1+\frac{l^2}{r^2}\right)^{-2N}}{r^3\left(l^2+r^2\right)^2}\left[l^4\left(2N^2\left(\Lambda r^3-3r+6M\right)\right.\right.\\
          & +\left.\left.N\left(3r-3\Lambda r^3\right)+\Lambda r^3\right)-l^2\left(Nr^2\left(\Lambda r^3+3r-6r_s\right)\right.\right.\\
          &-\left.\left.2\Lambda r^5\right)+\Lambda r^7\right]. 
    \end{split}
\end{equation}

We can easily verify from these expressions that when $r<<1$, the previous expressions reduce to Eqs. \eqref{Ricci approx} and \eqref{Kretshmann approx} as expected.

%%%%%%%%%%%%%%%%%%%%%%%%%%%%%%%%%%%%%%%%%%%%%%%%%%%%%%%%%%%%%%%%%%%%%%%%%%%%%%%%%

\section*{Acknowledgments}

The author thanks A. Herrera-Aguilar and D. Villaraos for fruitful discussions. The author also acknowledges financial assistance from CONAHCyT through the grant No. C-1325519.

%%%%%%%%%%%%%%%%%%%%%%%%%%%%%%%%%%%%%%%%%%%%%%%%%%%%%%%%%%%%%%%%%%%%%%%%%%%%%%%%%


\begin{thebibliography}{9}

\bibitem{EHT 1} K. Akiyama et al. (Event Horizon Telescope Collaboration), 
\textit{First M87 Event Horizon Telescope Results. IV. Imaging the Central
Supermassive Black Hole}, Astrophys. J. Lett. 875, L4 (2019).

\bibitem{EHT 2} K. Akiyama et al. (Event Horizon Telescope Collaboration), 
\textit{First Sagittarius A* Event Horizon Telescope Results. VI. Testing
the Black Hole Metric}, Astrophys. J. Lett. 930, L17 (2022).

\bibitem{LIGO 1} B. P. Abbott et al. (LIGO Scientific and Virgo Collabo-
rations), \textit{Observation of Gravitational Waves from a Bi-
nary Black Hole Merger}, Phys. Rev. Lett. 116, 061102
(2016).

\bibitem{LIGO 2} B. P. Abbott et al. (LIGO Scientific and Virgo Collabora-
tions), \textit{GW151226: Observation of Gravitational Waves
from a 22-Solar-Mass Binary Black Hole Coalescence},
Phys. Rev. Lett. 116, 241103 (2016).

\bibitem{Ghez 1} A. M. Ghez, S. Salim, N. N. Weinberg, J. R. Lu, T. Do,
J. K. Dunn, K. Matthews, M. R. Morris, S. Yelda, E. E.
Becklin, T. Kremenek, M. Milosavljevic and J. Naiman,
\textit{Measuring Distance and Properties of the Milky Way’s
Central Supermassive Black Hole with Stellar Orbits}, Astrophys. J. 689, 1044 (2008).

\bibitem{Ghez 2} M. R. Morris, L. Meyer and A. M. Ghez, \textit{Galactic center research: manifestations of the central black hole}, Res. Astron. Astrophys. 12, 995 (2012).

\bibitem{Genzel 1} A. Eckart and R. Genzel, \textit{Observations of stellar proper
motions near the Galactic Centre}, Nature (London) 383,
415 (1996).

\bibitem{Genzel 2} S. Gillessen, F. Eisenhauer, S. Trippe, T. Alexander, R.
Genzel, F. Martins and T. Ott, \textit{Monitoring stellar or-
bits around the massive black hole in the galactic center},
Astrophys. J. 692, 1075 (2009).

\bibitem{Stelle} K. S. Stelle, \textit{Renormalization of higher-derivative quantum gravity}, Phys. Rev. D 16, 953-969 (1977).

\bibitem{Modesto1} L. Modesto and L. Rachwal. \textit{Finite Conformal Quantum
Gravity and Nonsingular Spacetimes.} (2016).

\bibitem{Moffat1} J.W. Moffat, Phys. Rev. D41, 1177-1184 (1990).

\bibitem{Moffat2} B.J. Hand, J.W. Moffat, Phys. Rev. D43, 1896-1900
(1991).

\bibitem{Moffat3} D. Evens, J.W. Moffat, G. Kleppe, R.P. Woodard, Phys. Rev. D43, 499-519 (1991). 

\bibitem{Cornish1} N.J. Cornish, Mod.Phys.Lett.A7, 1895-1904 (1992). 

\bibitem{Cornish2} N.J. Cornish, Int. J. Mod. Phys. A7, 6121-6158 (1992).

\bibitem{Bambi1} Bambi C., Modesto L., Rachwal L., \textit{Spacetime completeness of nonsingular black holes in conformal gravity}, JCAP, 003, (2017). 

\bibitem{Bardeen} Bardeen, J., \textit{Non-singular general relativistic gravitational collapse}, presented at GR5, Tiflis, U.S.S.R., and published in the conference proceedings in the U.S.S.R. (1968).

\bibitem{Hayward} S. A. Hayward \textit{Formation and Evaporation of Nonsingular Black Holes}, Phys. Rev. Lett. 96, 031103 (2006). 

\bibitem{Beato-García1} E. Ayón-Beato and A. García, \textit{Regular Black Hole in General Relativity Coupled to Nonlinear Electrodynamics},
Phys. Rev. Lett. 80, 5056 (1998).

\bibitem{Beato-García2} E. Ayón-Beato and A. García, \textit{The Bardeen Model as a Nonlinear Magnetic Monopole},
Phys. Lett. B 493, 149 (2000).

\bibitem{NSSdS} D. A Easson, \textit{Nonsingular Schwarzschild–de Sitter black hole} Class. Quantum Grav. 35 235005 (2018).

\bibitem{Bambi3} C. Bambi, Z. Cao and L. Modesto, \textit{Testing conformal gravity with astrophysical black holes}. Phys. Rev. D 95, 064006 (2017).

\bibitem{Bambi4} M. Zhou, Z. Cao, A. Abdikamalov, D. Ayzenberg, C. Bambi, L. Modesto and S. Nampalliwar, \textit{Testing conformal gravity with the supermassive black hole in 1H707-495}. Phys. Rev. D 98, 024007 (2018).

\bibitem{Bambi5} M. Zhou, A. Abdikamalov, D. Ayzenberg, C. Bambi, L. Modesto, S. Nampalliwar and Y. Xu, \textit{Singularity-free black holes in conformal gravity: New observational constraints.} EPL 125, 30002 (2019).

\bibitem{Energy Conditions} Toshmatov, B., Bambi, C., Ahmedov, B. et al. \textit{Energy conditions of non-singular black hole spacetimes in conformal gravity}. Eur. Phys. J. C 77, 542 (2017).

\bibitem{Quasi-normal modes} C.Y. Chen, P. Chen., \textit{Gravitational perturbations of nonsingular black holes in conformal gravity}, Phys. Rev. D 99, 104003 (2019). 

\bibitem{HN} A. Herrera-Aguilar and U. Nucamendi, \textit{Kerr black hole parameters in terms of the redshift/blueshift of photons emitted by geodesic particles}, Phys. Rev. D 92, 045024 (2015). 

\bibitem{Martinez1} D. A. Martínez-Valera, M. Momennia and A. Herrera-Aguilar, \textit{Observational redshift from general spherically symmetric black holes}. Eur. Phys. J. C 84, 288 (2024).

\bibitem{Banerjee} P. Banerjee, A. Herrera-Aguilar, M. Momennia and U. Nucamendi, \textit{Mass and spin of Kerr black holes in terms of observational quantities: The dragging effect on the redshift}, Phys. Rev. D 105, 124037 (2022).

\bibitem{Martinez2} D. A. Martínez-Valera and A. Herrera-Aguilar, \textit{Parameter estimation of nonsingular black holes in conformal gravity using megamaser observational data from NGC 4258}, [arXiv:2504.04588 [gr-qc]], (2025). 

\bibitem{Villaraos} D. Villaraos, A. Herrera-Aguilar, U. Nucamendi, G.
González-Juárez and R. Lizardo-Castro, \textit{A general relativistic mass-to-distance ratio for a set of megamaser AGN black holes}, MNRAS 517,
4213 (2022).

\bibitem{Modesto4} L. Modesto. \textit{Super-renormalizable Multidimensional Quantum Gravity: Theory and Applications} ,Astron. Rev. 8.2 (2013).

\bibitem{Modesto1.1} Y. D. Li, L. Modesto, and L. Rachwał, \textit{Exact solutions and spacetime singularities in nonlocal gravity}. J. High Energ. Phys. 2015, 1–50 (2015). 

\bibitem{Modesto2} L. Modesto. \textit{Super-renormalizable quantum gravity}, Phys. Rev. D 86, 044005 (2012).

\bibitem{Modesto3} L. Modesto. \textit{Super-renormalizable Higher-Derivative Quantum Gravity}, [arXiv:1107.2403 [hep-th]] (2012).

\bibitem{Modesto5} Q. Li and L. Modesto. \textit{Galactic rotation curves in Einstein’s conformal gravity},  	[arXiv:1906.05185 [gr-qc]], (2019).

\bibitem{Modesto6} L. Modesto, L. Rachwał, and I.L. Shapiro, \textit{Renormalization group in super-renormalizable quantum gravity.} Eur. Phys. J. C 78, 555 (2018). 

\bibitem{Modesto7} L. Modesto \textit{Super-renormalizable or finite Lee–Wick quantum gravity}, Nuc. Phys. B, 909,  584-606, (2016).

\bibitem{Modesto8} L. Modesto, \textit{Finite quantum gravity.} [arXiv:1305.6741  [hep-th]], (2013).

\bibitem{Modesto9} T. Zhou, and L. Modesto, \textit{Geodesic incompleteness of some popular regular black holes.} Phys. Rev. D 107, 044016, (2023).

\bibitem{UniversalFQG}  L. Modesto, and L. Rachwał, \textit{Universally finite gravitational and gauge theories}, Nuc. Phys. B, 900,  147-169, (2015).

\bibitem{Tomboulis} E. T. Tomboulis, \textit{Superrenormalizable gauge and gravitational theories}, [arXiv: 9702146  [hep-th]], (1997).

\bibitem{Theorems} M. A. Luty, J. Polchinski and R. Rattazzi, \textit{The $\alpha$-theorem and the Asymptotics of 4D Quantum Field Theory}, JHEP 1301, 152 (2013).

\bibitem{Bambi2} Q. Zhang, , L. Modesto,  and C. Bambi, \textit{A general study of regular and singular black hole solutions in Einstein’s conformal gravity.} Eur. Phys. J. C 78, 506, (2018).

\bibitem{2mass} Uzan, JP., Ellis, G.F.R., Larena, J. A. \textit{A two-mass expanding exact space-time solution} Gen Relativ. Gravit. 43, 191–205 (2011). https://doi.org/10.1007/s10714-010-1081-6






\end{thebibliography}
\end{document}